\documentclass{article}

\usepackage{arxiv}

\usepackage[utf8]{inputenc} % allow utf-8 input
\usepackage[T1]{fontenc}    % use 8-bit T1 fonts
\usepackage{hyperref}       % hyperlinks
\usepackage{url}            % simple URL typesetting
\usepackage{booktabs}       % professional-quality tables
\usepackage{amsfonts}       % blackboard math symbols
\usepackage{amssymb}        % needed for \leqslant
\usepackage{amsmath}        % needed for cases
\usepackage{nicefrac}       % compact symbols for 1/2, etc.
\usepackage{microtype}      % microtypography
\usepackage{lipsum}		% Can be removed after putting your text content
\usepackage{graphicx}
\usepackage[numbers]{natbib}
\usepackage{doi}
\usepackage{wrapfig}
\usepackage{tikz-cd}

\title{The free group in R}

\author{ \href{https://orcid.org/0000-0001-5982-0415}{\includegraphics[width=0.03\textwidth]{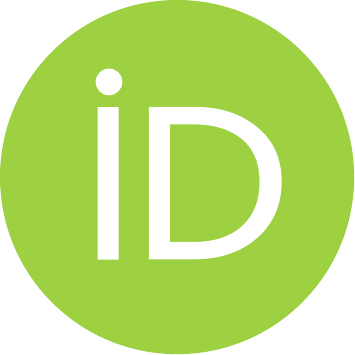}\hspace{1mm}Robin K. S.~Hankin}\thanks{\href{https://academics.aut.ac.nz/robin.hankin}{work};  
\href{https://www.youtube.com/watch?v=JzCX3FqDIOc&list=PL9_n3Tqzq9iWtgD8POJFdnVUCZ_zw6OiB&ab_channel=TrinTragulaGeneralRelativity}{play}} \\
 Auckland University of Technology\\
	\texttt{hankin.robin@gmail.com} \\
}

\renewcommand{\shorttitle}{The free group in R}

\hypersetup{
pdftitle={The free group in R},
pdfsubject={q-bio.NC, q-bio.QM},
pdfauthor={Robin K. S.~Hankin},
pdfkeywords={the free group}
}

\usepackage{Sweave}
\begin{document}
\maketitle

\begin{abstract}
Here I present the {\tt freegroup} package for working with the free
group on a finite set of symbols.  The package is vectorised;
internally it uses an efficient matrix-based representation for free
group objects but uses a configurable print method.  A range of
R-centric functionality is provided.  It is available on CRAN at
\url{https://CRAN.R-project.org/package=freegroup}.
\end{abstract}

\section{Introduction}

\setlength{\intextsep}{0pt}
\begin{wrapfigure}{r}{0.2\textwidth}
  \begin{center}
\includegraphics[width=1in]{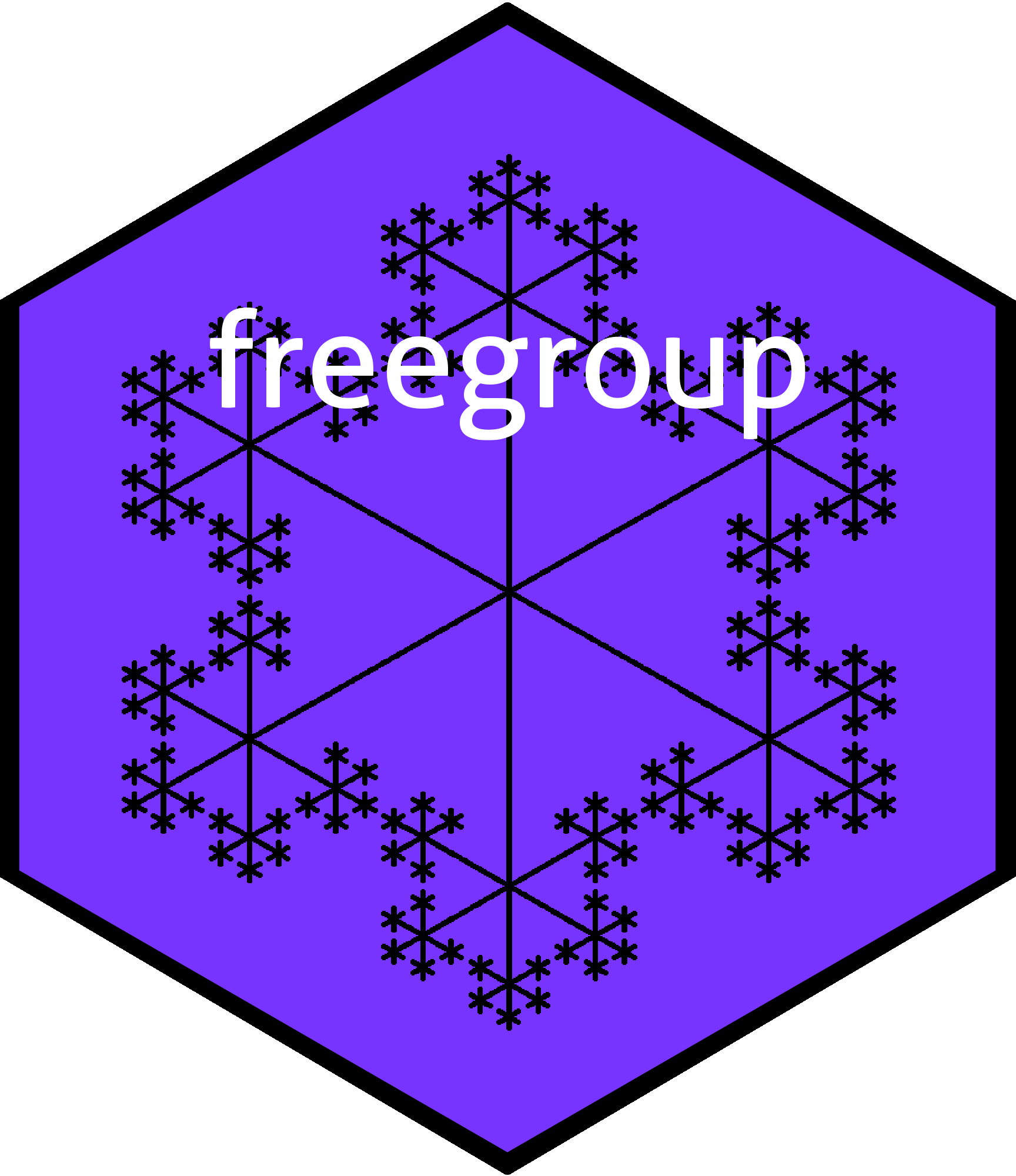}
  \end{center}
\end{wrapfigure}

The free group is an interesting and instructive mathematical object
with a rich structure that illustrates many concepts of elementary
group theory.  The {\tt freegroup} package provides some functionality
for manipulating the free group on a finite list of symbols.  It is
written in the R programming language~\cite{rcore2022} and is
available on CRAN at
\url{https://CRAN.R-project.org/package=freegroup}.

Informally, the {\em free group} $\left(X,\circ\right)$ on a set
$S=\{a,b,c,\ldots,z\}$ is the set $X$ of {\em words} that are objects
like $W=c^{-4}bb^2aa^{-1}ca$, with a group operation of string
juxtaposition.  Usually one works only with words that are in
``reduced form'', which has successive powers of the same symbol
combined, so $W$ would be equal to $c^{-4}b^3ca$; see how $b$ appears
to the third power and the $a$ term in the middle has vanished through
cancellation.  The group operation of juxtaposition is formally
indicated by $\circ$, but this is often omitted in algebraic notation;
thus, for example $a^2b^{-3}c^2\circ c^{-2}ba =a^2b^{-3}c^2c^{-2}ba
=a^2b^{-2}ba$.

\paragraph{Formal definition}
If $X$ is a set, then a group $F$ is called {\em the free group on
$X$} if there is a set map $\Psi\colon X\longrightarrow F$, and for
any group $G$ and set map $\Phi\colon X\longrightarrow G$, there is a
unique homomorphism $\alpha\colon F\longrightarrow G$ such that
$\alpha\circ\Psi=\Phi$, that is, the diagram below commutes:

\begin{center}
\begin{tikzcd}
X \arrow[r,"\Psi"] \arrow[dr,"\Phi"]
& F \arrow[d,"\alpha"]\\
& G
\end{tikzcd}
\end{center}

It can be shown that $F$ is unique up to group isomorphism; every
group is a quotient of a free group.

\subsection{Existing work}

Computational support for working with the free group is provided as
part of a number of algebra systems including GAP~\cite{GAP4},
Sage~\citep{sagemath2019}, and {\tt sympy}~\citep{sympy2017} although
in those systems the emphasis is on finitely presented groups, not in
scope for the {\tt freegroup} package.  There are also a number of
closed-source proprietary systems.

\section{The package in use}

In the {\tt freegroup} package, a word is represented by a two-row
integer matrix; the top row is the integer representation of the
symbol and the second row is the corresponding power.  For example, to
represent $a^2b^{-3}ac^2a^{-2}$ we would identify $a$ as 1, $b$ as 2,
etc and write

\begin{Schunk}
\begin{Sinput}
> (M <- rbind(c(1,2,3,3,1),c(2,-3,2,3,-2)))
\end{Sinput}
\begin{Soutput}
     [,1] [,2] [,3] [,4] [,5]
[1,]    1    2    3    3    1
[2,]    2   -3    2    3   -2
\end{Soutput}
\end{Schunk}

(see how negative entries in the second row correspond to negative
powers).  Then to convert to a more useful form we would have

\begin{Schunk}
\begin{Sinput}
> library("freegroup")
> (x <- free(M))
\end{Sinput}
\begin{Soutput}
[1] a^2.b^-3.c^5.a^-2
\end{Soutput}
\end{Schunk}

The representation for {\tt R} object {\tt x} is still a two-row
matrix, but the print method is active and uses a more visually
appealing scheme.  The default alphabet used is {\tt letters}.  On
this understanding, we can coerce strings to free objects:

\begin{Schunk}
\begin{Sinput}
> (y <- as.free("aabbbcccc"))
\end{Sinput}
\begin{Soutput}
[1] a^2.b^3.c^4
\end{Soutput}
\end{Schunk}

The free group operation is simply juxtaposition, represented here by
the plus symbol:

\begin{Schunk}
\begin{Sinput}
> x+y
\end{Sinput}
\begin{Soutput}
[1] a^2.b^-3.c^5.b^3.c^4
\end{Soutput}
\end{Schunk}

(see how the $a$ ``cancels out'' in the juxtaposition).  One motivation
for the use of ``{\tt +}'' rather than ``{\tt *}'' is that
{\tt Python} uses ``{\tt +}'' for appending strings:

\begin{verbatim}
>>> "a" + "abc"
'aabc'
>>> 
\end{verbatim}

However, note that in mathematics the ``{\tt +}'' symbol is usually
reserved for commutative and associative operations; string
juxtaposition is associative but not commutative.  Multiplication by
integers---denoted in {\tt freegroup} idiom by ``{\tt *}''---is also
defined.  Suppose we want to concatenate 3 copies of {\tt x}:

\begin{Schunk}
\begin{Sinput}
> x
\end{Sinput}
\begin{Soutput}
[1] a^2.b^-3.c^5.a^-2
\end{Soutput}
\begin{Sinput}
> x*3
\end{Sinput}
\begin{Soutput}
[1] a^2.b^-3.c^5.b^-3.c^5.b^-3.c^5.a^-2
\end{Soutput}
\end{Schunk}

However, with these definitions the distributive law is broken: in
general {\tt n*(a + b)} is not equal to {\tt n*a + n*b}.  
\begin{Schunk}
\begin{Sinput}
> (a <- as.free("aab"))
\end{Sinput}
\begin{Soutput}
[1] a^2.b
\end{Soutput}
\begin{Sinput}
> (b <- abc(1:4))
\end{Sinput}
\begin{Soutput}
[1] a       a.b     a.b.c   a.b.c.d
\end{Soutput}
\begin{Sinput}
> 2*(a+b)
\end{Sinput}
\begin{Soutput}
[1] a^2.b.a^3.b.a               a^2.b.a.b.a^2.b.a.b        
[3] a^2.b.a.b.c.a^2.b.a.b.c     a^2.b.a.b.c.d.a^2.b.a.b.c.d
\end{Soutput}
\begin{Sinput}
> 2*a + 2*b
\end{Sinput}
\begin{Soutput}
[1] a^2.b.a^2.b.a^2             a^2.b.a^2.b.a.b.a.b        
[3] a^2.b.a^2.b.a.b.c.a.b.c     a^2.b.a^2.b.a.b.c.d.a.b.c.d
\end{Soutput}
\end{Schunk}

However, the abelianizations are equal:

\begin{Schunk}
\begin{Sinput}
> abelianize(2*(a+b)) == abelianize(2*a + 2*b)
\end{Sinput}
\begin{Soutput}
[1] TRUE TRUE TRUE TRUE
\end{Soutput}
\end{Schunk}

The package is vectorized and is generally consistent with R indexing
mechanisms:

\begin{Schunk}
\begin{Sinput}
> x*(0:3)
\end{Sinput}
\begin{Soutput}
[1] 0                                   a^2.b^-3.c^5.a^-2                  
[3] a^2.b^-3.c^5.b^-3.c^5.a^-2          a^2.b^-3.c^5.b^-3.c^5.b^-3.c^5.a^-2
\end{Soutput}
\end{Schunk}

There are a few methods for creating free objects, for example:

\begin{Schunk}
\begin{Sinput}
> alpha(1:9)
\end{Sinput}
\begin{Soutput}
[1] a b c d e f g h i
\end{Soutput}
\begin{Sinput}
> abc(1:9)
\end{Sinput}
\begin{Soutput}
[1] a                 a.b               a.b.c             a.b.c.d          
[5] a.b.c.d.e         a.b.c.d.e.f       a.b.c.d.e.f.g     a.b.c.d.e.f.g.h  
[9] a.b.c.d.e.f.g.h.i
\end{Soutput}
\end{Schunk}

And we can also generate random free objects:

\begin{Schunk}
\begin{Sinput}
> rfree(10,4)
\end{Sinput}
\begin{Soutput}
 [1] a^3.b^2.d^4.c^3 c^-4.b^4        d^3.b^-4        c^-4           
 [5] a^3.d.c^-2.b^-1 b^-3.a^3.b^-2   a^-2.b^-5.d     c.a^-2.d.a^-1  
 [9] b^-1.c          b^-3.d^-4.b^-2 
\end{Soutput}
\end{Schunk}

Inverses are calculated using unary or binary minus:

\begin{Schunk}
\begin{Sinput}
> (u <- rfree(10,4))
\end{Sinput}
\begin{Soutput}
 [1] a^-2.b^7         c^-2.a^3.c^2     d^-3.c^5         a               
 [5] b^3.c^-2.a^4     a^-2.d^4.a^7     d^2.c^-1.d^-3    d^-4.c^-3.d.c^-1
 [9] c^-4.b^2.c^2     b^-3.d^-1       
\end{Soutput}
\begin{Sinput}
> -u
\end{Sinput}
\begin{Soutput}
 [1] b^-7.a^2       c^-2.a^-3.c^2  c^-5.d^3       a^-1           a^-4.c^2.b^-3 
 [6] a^-7.d^-4.a^2  d^3.c.d^-2     c.d^-1.c^3.d^4 c^-2.b^-2.c^4  d.b^3         
\end{Soutput}
\begin{Sinput}
> u-u
\end{Sinput}
\begin{Soutput}
 [1] 0 0 0 0 0 0 0 0 0 0
\end{Soutput}
\end{Schunk}

Above, we see the print method has special dispensation for the
identity element, displaying it as {\tt 0}.  We can take the ``sum''
of a vector of free objects simply by juxtaposing the elements:

\begin{Schunk}
\begin{Sinput}
> sum(u)
\end{Sinput}
\begin{Soutput}
[1] a^-2.b^7.c^-2.a^3.c^2.d^-3.c^5.a.b^3.c^-2.a^2.d^4.a^7.d^2.c^-1.d^-7.c^-3.d.c^-5.b^2.c^2.b^-3.d^-1
\end{Soutput}
\end{Schunk}

Powers are defined as per group conjugation: $x^y=y^{-1}xy$, {\tt
x\string^y=y\string^{-1}xy} (or, written in additive notation, {\tt
-y+x+y}).  The nomenclature is motivated by the identities $x^{yz} =
\left(x^y\right)^z$ and $(xy)^z=x^zy^z$.  In multiplicative notation
this would translate to {\tt x\string^(yz)==(x\string^y)\string^z} and
{\tt (xy)\string^z=(x\string^z)*(y\string^z)}.  Additive notation is
somewhat less appealing and the identities appear as {\tt
x\string^(y+z)==(x\string^y)\string^z} and {\tt
(x+y)\string^z=(x\string^z)+(y\string^z)}.  Numerical verification
follows:

\begin{Schunk}
\begin{Sinput}
> u
\end{Sinput}
\begin{Soutput}
 [1] a^-2.b^7         c^-2.a^3.c^2     d^-3.c^5         a               
 [5] b^3.c^-2.a^4     a^-2.d^4.a^7     d^2.c^-1.d^-3    d^-4.c^-3.d.c^-1
 [9] c^-4.b^2.c^2     b^-3.d^-1       
\end{Soutput}
\begin{Sinput}
> y <- alpha(25)
> z <- alpha(26)
> all(c(u^(y+z) == (u^y)^z, (u+y)^z == u^z + y^z))
\end{Sinput}
\begin{Soutput}
[1] TRUE
\end{Soutput}
\end{Schunk}

We may generalize the second identity to arbitrary concatenation; thus:

\begin{Schunk}
\begin{Sinput}
> sum(u^z) == sum(u)^z
\end{Sinput}
\begin{Soutput}
[1] TRUE
\end{Soutput}
\end{Schunk}

\section{Commutator brackets and the Hall-Witt identity}

The package also includes a commutator bracket, defined as
$[x,y]=x^{-1}y^{-1}xy$ or in package idiom {\tt .[x,y]=-x-y+x+y}:
 
\begin{Schunk}
\begin{Sinput}
> .[y,z]
\end{Sinput}
\begin{Soutput}
[1] y^-1.z^-1.y.z
\end{Soutput}
\end{Schunk}

It is worth observing that the commutator bracket as defined does not
obey the Jacobi relation$ [x,[y,z]] + [y,[z,x]] +[z,[x,y]]= 0$:

\begin{Schunk}
\begin{Sinput}
> x <- rfree()
> y <- rfree()
> z <- rfree()
> is.id(.[x,.[y,z]] + .[y,.[z,x]] + .[z,.[x,y]])
\end{Sinput}
\begin{Soutput}
[1] FALSE FALSE FALSE FALSE FALSE FALSE FALSE
\end{Soutput}
\end{Schunk}

However, the Hall-Witt identity $[[x,-y],z]^y + [[y,-z],x]^z
+[[z,-x],y]^x$ is satisfied:

\begin{Schunk}
\begin{Sinput}
> is.id(.[.[x,-y],z]^y + .[.[y,-z],x]^z + .[.[z,-x],y]^x)
\end{Sinput}
\begin{Soutput}
[1] TRUE TRUE TRUE TRUE TRUE TRUE TRUE
\end{Soutput}
\end{Schunk}

\section{Print method}

If we have more than 26 symbols the print method runs out of letters:

\begin{Schunk}
\begin{Sinput}
> alpha(1:30)
\end{Sinput}
\begin{Soutput}
 [1] a  b  c  d  e  f  g  h  i  j  k  l  m  n  o  p  q  r  s  t  u  v  w  x  y 
[26] z  NA NA NA NA
\end{Soutput}
\end{Schunk}

If this is a problem (it might not be: the print method might not be
important) it is possible to override the default symbol set:

\begin{Schunk}
\begin{Sinput}
> options(symbols = state.abb)
> alpha(1:30)
\end{Sinput}
\begin{Soutput}
 [1] AL AK AZ AR CA CO CT DE FL GA HI ID IL IN IA KS KY LA ME MD MA MI MN MS MO
[26] MT NE NV NH NJ
\end{Soutput}
\end{Schunk}

\section{Conclusions and further work}

The {\tt freegroup} package furnishes a consistent and documented
suite of reasonably efficient {\tt R}-centric functionality.  Further
work might include the efficient detection and generation of
square-free words.

\bibliographystyle{apalike}
\bibliography{freegroup_arxiv}

\end{document}